\documentclass[aps,prb,10pt,twocolumn,superscriptaddress]{revtex4}
\pdfoutput=1
\usepackage{graphicx}
\usepackage{amssymb}
\usepackage{amsmath}

\def\cm{cm$^{-1}$} 
\begin{document}

\title{Direct evidence of overdamped Peierls-coupled modes\\ in TTF-CA
temperature-induced phase transition}

\author{A. Girlando}
\author{M. Masino}
\author{A. Painelli}
\affiliation{Dipartimento di Chimica Generale ed Inorganica, Chimica
Analitica e Chimica Fisica, and INSTM-UdR Parma, Universit\`a di Parma, Parco Area
delle Scienze, 43100-I Parma, Italy}
\author{N. Drichko}
\affiliation{1.~Physikalisches Institut, Universit{\"a}t
Stuttgart, Pfaffenwaldring 57, 70550 Stuttgart, Germany}
\affiliation{Ioffe Physico-Technical Institute Russian Academy of
Sciences, Politeknicheskaya 26, 194021 St. Petersburg, Russia}
\author{M. Dressel}
\affiliation{1.~Physikalisches Institut, Universit{\"a}t
Stuttgart, Pfaffenwaldring 57, 70550 Stuttgart, Germany}
\author{A. Brillante}\author{R. G. Della Valle}\author{E. Venuti}
\affiliation{Dipartimento di Chimica Fisica e Inorganica, and INSTM-UdR Bologna,
Universit\`a di Bologna, Viale Risorgimento 4, 40136-I Bologna, Italy}


\begin{abstract}
In this paper we elucidate the optical response resulting from the
interplay of charge distribution (ionicity) and Peierls
instability (dimerization) in the neutral-ionic, ferroelectric
phase transition of tetrathiafulvalene-chloranil (TTF-CA), a
mixed-stack quasi-one-dimensional charge-transfer crystal. We
present far-infrared reflectivity measurements down to $5$~\cm\ as
a function of temperature above the phase transition (300 -- 82
K). The coupling between electrons and lattice phonons in the
pre-transitional regime is analyzed on the basis of phonon
eigenvectors and polarizability calculations of the
one-dimensional Peierls-Hubbard model. We find a multi-phonon
Peierls coupling, but on approaching the transition the spectral
weight and the coupling shift progressively towards the phonons at
lower frequencies, resulting in a soft-mode behavior only for the
lowest frequency phonon near the transition temperature. Moreover, in the
proximity of the phase transition, the lowest-frequency phonon becomes
overdamped, due to anharmonicity induced by its coupling to
electrons. The implications of these findings for the
neutral-ionic transition mechanism is shortly discussed.
\end{abstract}
\maketitle

\section{Introduction}
\label{sec:introduction}

In a formulation similar to a student's exercise, more than fifty
year ago Rudolf Peierls pointed out that one-dimensional (1D)
metals are intrinsically unstable with respect to lattice
distortions opening a gap at the Fermi energy.\cite{peierls} It
did not take long to realize that the quasi-1D electronic
structure of organic charge-transfer (CT) conductors offers an
almost ideal model for the Peierls mechanism. Extensive
investigations in the 1970s revealed that the Peierls instability
may occur also in 1D Mott insulators, due to coupling of the
lattice to spin (spin-Peierls),\cite{mcconnell} or to both
electronic and spin degrees of freedom (``generalized" Peierls
instability).\cite{bray} At the time, the attention was almost
exclusively devoted to segregated-stack CT crystals, where
identical electron-donor (D) or acceptor (A) $\pi$-molecules are
arranged face-to-face along one direction. Mixed stacking
(...DADA...) received far less attention, at least until a new
kind of phase transition was discovered in these systems: the
so-called neutral-ionic phase transition (NIT).\cite{nitP,nitT} It
is characterized by a change in the ionicity $\varrho$, i.e.\ the
charge transfer from D to A in the ground state. Temperature $T$
or pressure $p$ may induce a change from a quasi-neutral
\textit{N}, $\varrho \lesssim $ 0.5) to a quasi-ionic (\textit{I},
$\varrho \gtrsim $ 0.5) ground state. It was early recognized that
the ionic state is subject to the Peierls instability, yielding a
stack dimerization (..DA DA DA..).\cite{GP86,nagaosa86}

NIT are complex phenomena, resulting from the interplay between
two order parameters, the ionicity $\varrho$ and the dimerization
$\delta$, and present a rich and intriguing
phenomenology.\cite{ricemem,tokurareview} Some of the phenomena
occurring on approaching the phase transition,
like the dramatic increase of the dielectric constant or the
observation of diffuse scattering signals in x-ray
diffraction,\cite{TTFCAdiel,buron06} have been recently
ascribed to charge oscillations associated with the soft mode that
yields the stack dimerization, namely, the Peierls
mode.\cite{freo02,soos04,davino07} It is then of fundamental
importance to identify the Peierls mode in order to confirm the
above suggestion and to shed light on the NIT mechanism. At
variance with segregated stack crystals, the Peierls mode is
optically active in mixed stack structures,\cite{GP86} which
offers a rather unique possibility for its optical detection and
characterization. Finally, since the dimerized ionic stack is
potentially ferroelectric,\cite{tokura89} the Peierls mode is the
counterpart of the so-called ferroelectric mode of perovskite-type
crystals.\cite{nakamura66}

Direct and clear experimental identification and characterization
of the Peierls mode and of its role in the NIT has proved to be
more difficult than expected.\cite{moreac96,okimoto01} One of the
reasons is that the phonon structure of molecular crystals is 
very complex, and several modes are likely coupled to CT
electrons. Therefore the analysis of the spectra requires a
detailed and reliable understanding of the lattice-phonon
dynamics. We have recently undertaken such analysis for
tetrathiafulvalene-chloranil (TTF-CA), which undergoes a
first-order temperature induced NIT at 81 K.\cite{ricemem} In a
first paper (hereafter Paper I)\cite{paperI} we presented a quasi
harmonic lattice dynamics (QHLD) calculations of TTF-CA lattice
phonons above and below the transition temperature, and gave a
first partial interpretation of the infrared (IR) and Raman
spectra.

The specific goal and achievement of the present paper is the
identification of the Peierls (soft) modes in TTF-CA neutral
phase. We present detailed and improved data for single crystals
of TTF-CA in the spectral region $5-200$~\cm\ collected in the
temperature range down to 82 K. Particular attention is devoted to
the submillimeter spectral range, below 30~\cm. The experimental
data are analyzed in terms of a multi-mode Peierls model, that
combines the information on lattice phonon dynamics reported in
Paper I,\cite{paperI} with that on the electronic structure as obtained
from the exact diagonalization of the 1D Peierls-Hubbard
Hamiltonian in Ref.~\onlinecite{soos04}. The model fully accounts
for the experimental data and allows us to demonstrate that
\textit{mode mixing} and \textit{overdamping} are the key to
explain the changes of the far-IR spectra on approaching the
transition. The effective Peierls mode detected by an analysis of
the combination modes in the mid-IR region\cite{sidebands} is also
accounted for. In addition, the dielectric constant anomaly at the
transition\cite{TTFCAdiel} is quantitatively reproduced.

\section{Experimental methods}
\label{sec:experimental}

TTF-CA single crystals were grown by sublimation. Crystals as
large as $4~{\rm mm}\times 4~{\rm mm}$ allowed for measurement
down to 5~\cm. The reflectivity was probed on well-developed
naturally grown $ab$ crystal faces along the optical axes. The
crystals were oriented at room temperature according to the known
spectra in mid-IR region, using a Hyperion IR
microscope attached to Bruker IFS66v Fourier-transform
spectrometer. In this paper we present the spectra polarized
parallel to the $a$ stack axis; only these spectra are relevant to
the analysis of Peierls coupling in the neutral
phase.\cite{paperI}

The reflectivity spectra were collected by three different setups
and techniques, depending on the spectral region of interest:
(i)~In the $5-30$~\cm\ region, we employed a quasi-optical
submillimeter spectrometer, equipped with backward wave oscillator
as coherent and tunable source; a Golay-cell served as a detector.
Temperature dependent measurements were conducted in an
exchange-gas helium cryostat. The absolute reflectivity was
obtained by comparing sample reflectance to reflectance of an
aluminum mirror fixed on the same aperture. The spectral
resolution in this range is better than 0.01~\cm. (ii)~In the
$20-150$~\cm\ range the reflectivity was measured by a Bruker
IFS66v with a spectral resolution of 0.5~\cm. The spectrometer was
equipped with 1.4~K Si bolometer and a CryoVac cold-finger
cryostat. The sample was fixed on the cold finger by carbon paste;
a good temperature contact was ensured by Apiezon paste. To get
absolute values of reflectivity, the {\it in-situ}
gold-evaporation technique was adopted:\cite{homes93} after
measuring the temperature dependence of the sample reflectivity,
gold is evaporated onto the sample surface, and used as reference.
(iii)~Reflectance measurements in the $100-6000$~\cm\ range were
done in a Bruker IFS 113v FTIR spectrometer equipped with a 4.2~K
Si bolometer, an MCT detector, and an exchange-gas helium
cryostat. The absolute reflectivity values of the sample have been
obtained by a comparison to a Au mirror of the same size. In this
spectral range a resolution of 2~\cm\ was chosen. The reflectivity
values agree well with available literature data,\cite{jacobsen83}
which extend from 3000 to 14\,000~\cm\ and thus cover the charge
transfer transition. The extension to high wavenumbers is needed
to obtain reliable Kramers-Kronig transformation (KKT) of the data.

We ensured that good overlap exists between the data measured in
the three spectral regions. In order to match the absolute values,
the submillimeter data were rescaled to those obtained by the
Bruker IFS66v FTIR because the {\it in-situ} gold-evaporation
technique ensures the most precise measurements of the absolute
reflectivity and does not depend on the quality of the sample
surface.

In the KKT analysis the data were extrapolated below
5~\cm\ to fit the known values of the static dielectric constant
$\epsilon_1$.\cite{TTFCAdiel} We also verified that an
extrapolation to a constant value does not significantly affect
the portion of the spectra for which experimental data exist. At
frequencies above 14\,000~\cm\ an extrapolation $R(\omega)\propto
\omega^{-2}$ has been used.

\section{Reflectivity spectra}
\label{sec:expspectra}
\subsection{Spectral data}
\label{subsec:reflspectra}
\begin{figure}[ht]
 \centering
 \includegraphics[width=8.5cm]{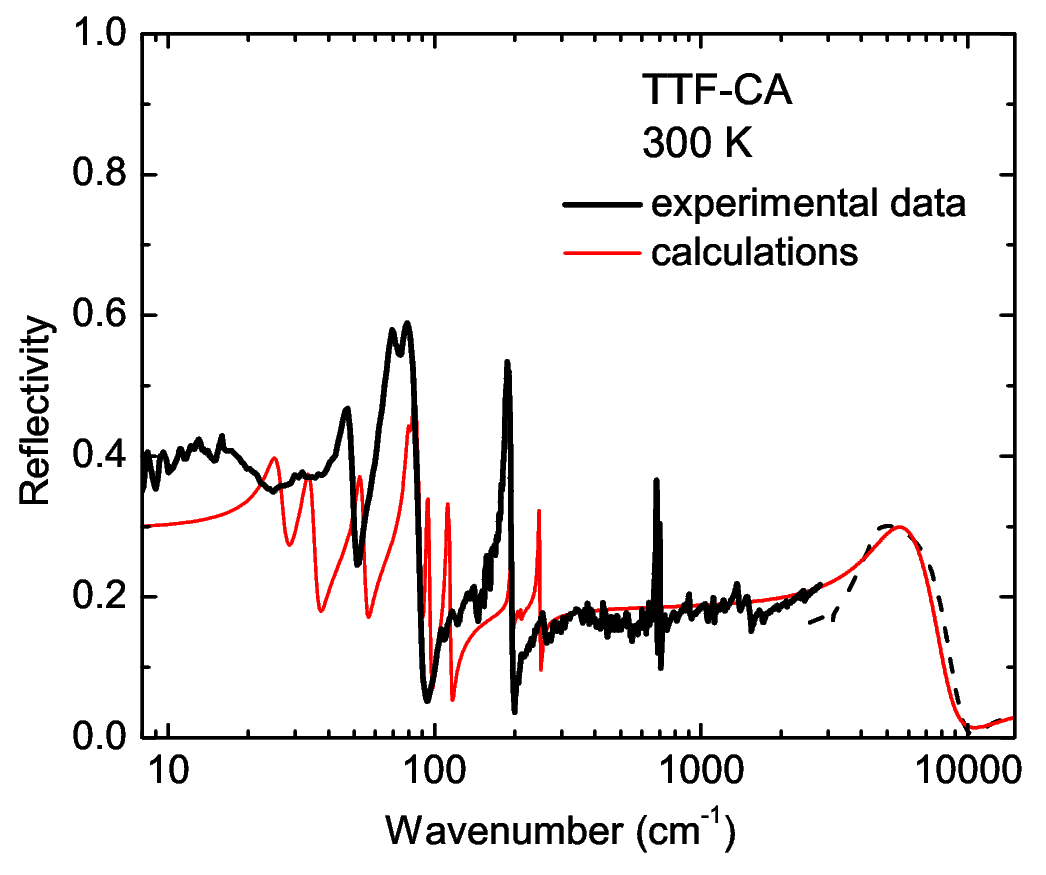}
  \caption{(color online) Experimental and calculated
(Section \ref{sec:modelling}) reflectivity spectra of TTF-CA at
$T=300$~K, with the polarization $E\parallel a$. The spectral
region above 3000~\cm\ (dashed curve) is taken from
Ref.~\onlinecite{jacobsen83}. Notice the logarithmic frequency
scale.}
 \label{fig:allRefl_300K}
\end{figure}

Fig. \ref{fig:allRefl_300K} shows the room temperature TTF-CA
reflectivity along the stack axis for the whole measured spectral
range, with extension to 14\,000~\cm\ taken from
Ref.~\onlinecite{jacobsen83}. The reflectivity is rather low, as
expected for a non-metallic compound. The charge transfer
electronic transition occurs around 5000~\cm\, and at lower
frequencies reflectivity gets down to about 0.2, with no relevant
feature observed until we encounter the highest frequency
out-of-plane intramolecular modes, at 701 and 682~\cm,
corresponding to CA and TTF, respectively.\cite{girlando83} Below
250~\cm\ we find a couple of other modes of prevailing
out-of-plane, intramolecular character, and finally the
intermolecular modes we want to discuss in detail in the present
paper.
\begin{figure}[hb]
\includegraphics[width=8.5cm]{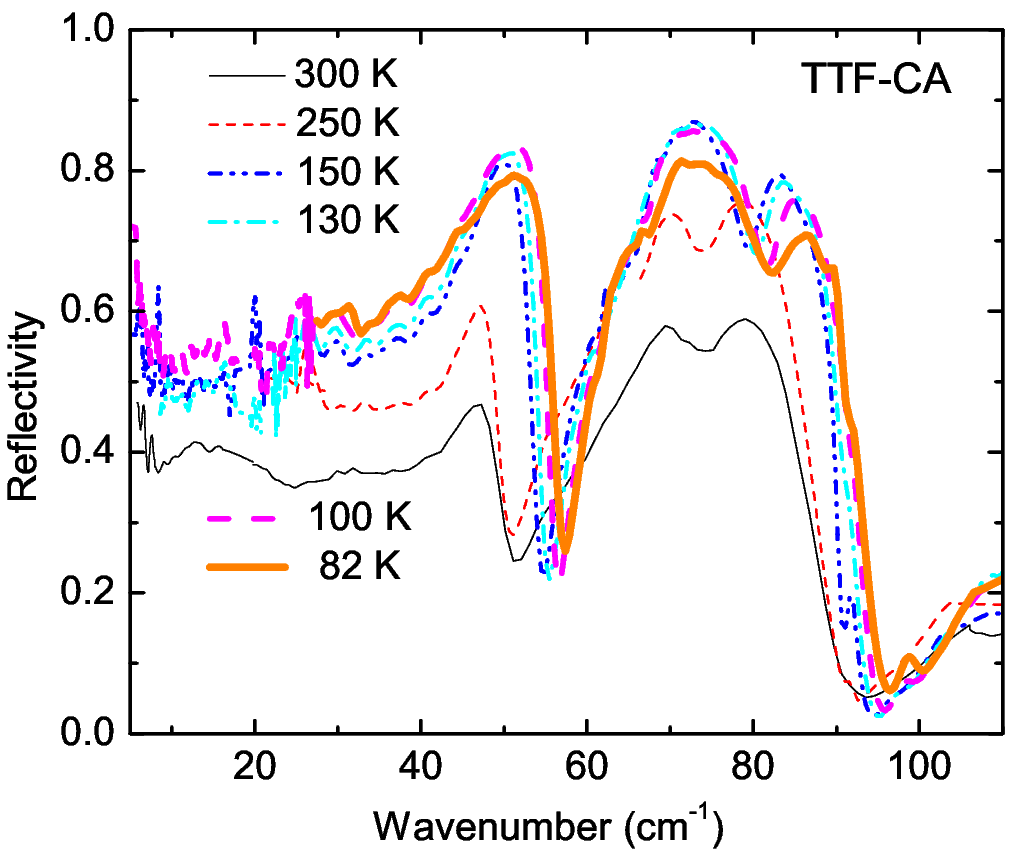}
\caption{(color online) Reflectivity spectra of TTF-CA single
crystal along the $a$ axis taken at
different temperatures $T > T_c$ above the {\it N -- I} transition.}
\label{fig:reflect_T}
\end{figure}

The temperature dependence of the low-frequency ($5-120$~\cm)
reflectivity spectra is presented in Fig.~\ref{fig:reflect_T}. At
room temperature a group of strong phonon bands is observed just
below 100~\cm. At lower frequencies the reflectivity remains
approximately at the level of 0.4 down to the lowest measured data
(5~\cm~). On cooling the sample from 300 to 82~K, the reflectivity
above 120~\cm\ reveals only unimportant changes, and is therefore
not reported. Instead Fig.~\ref{fig:reflect_T} shows that dramatic
changes occur below 100~\cm, where reflectivity rises by 1.5 times
on lowering temperature to 82~K, just before the NIT. As
demonstrated in Paper I,\cite{paperI} below the transition
temperature ($T_c$ = 81 K) the reflectivity drops down to values
below those at $T=300$~K, and does not change considerably on
further cooling to 10 K.
\begin{figure}
\includegraphics[width=8.5cm]{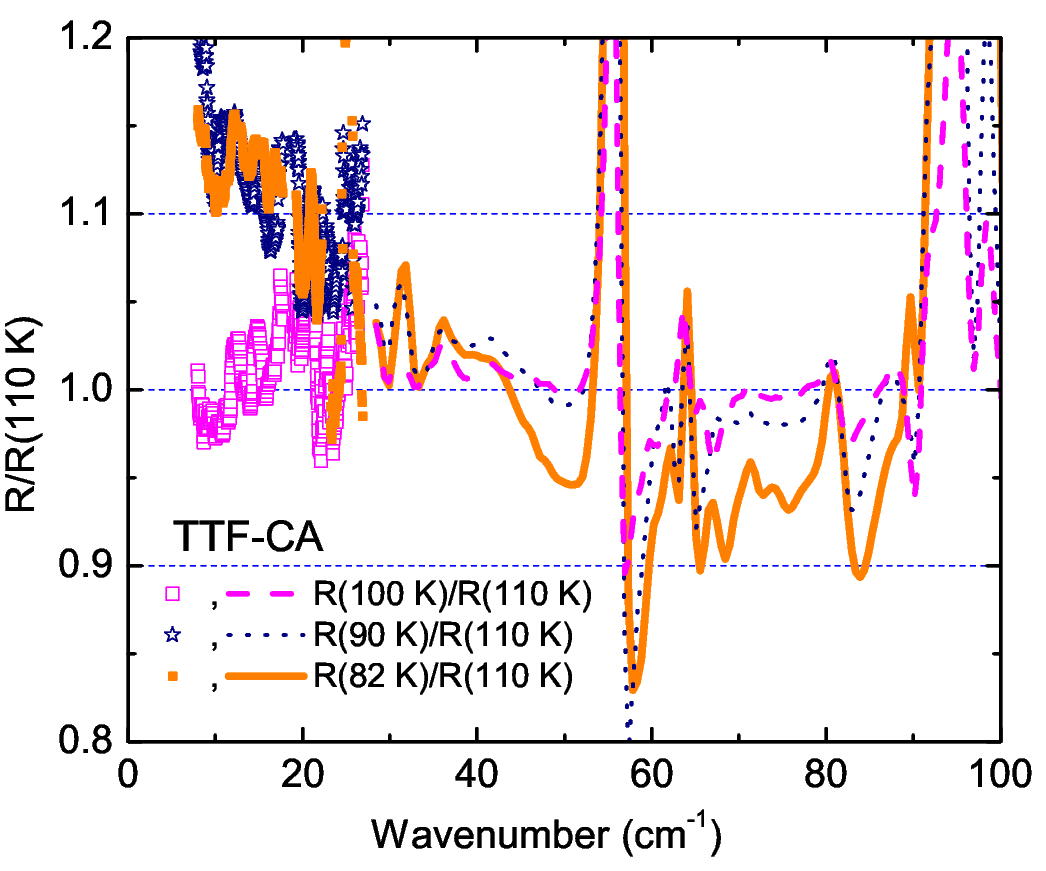}
  \caption{(color online) Ratio of the of the reflectivity spectra at 82, 90, and 100~K
   to the reflectivity at 110 K. Measurements in the range below 25 \cm\
   performed by using the submillimeter spectrometer (points),
   data above 25 \cm\ collected by using Bruker 66v spectrometer
   (lines). Note the increase of reflectance below 25~\cm\ at
   temperatures close to the {\it N -- I} transition temperature.}
\label{fig:dividing}
\end{figure}

A more detailed inspection of Fig.~\ref{fig:reflect_T} provides
evidence that the strongest increase of reflectivity, from about
0.4 to 0.6, occurs on cooling between 300 and 150~K. Some increase
of reflectivity below approximately 40~\cm\ occurs on further
cooling, indicating a shift of spectral weight to lower
frequencies. The quality of the data in the low-frequency region,
essential to appreciate the shift of spectral weight, is
illustrated in Fig.~\ref{fig:dividing}, where the reflectivity at
$T$ = 82, 90 and 100 K is normalized to the reflectivity at 110~K.
This figure illustrate the good overlap between 
submillimeter and FIR data, collected with two different
instrumentations (see Section \ref{sec:experimental}). The
reflectivity ratio $R(T)/R(110~{\rm K})$ for these two spectral
regions have indeed differences smaller than the noise of the
measurements. Thus we safely confirm that by lowering $T$ below
100 K the reflectivity drops in the range 30-100~\cm\ range and
increases below, with a redistribution of the overall spectral
weight.

\subsection{Optical conductivity}
\label{subsec:condspectra}

\begin{figure}
\includegraphics[width=8.5cm]{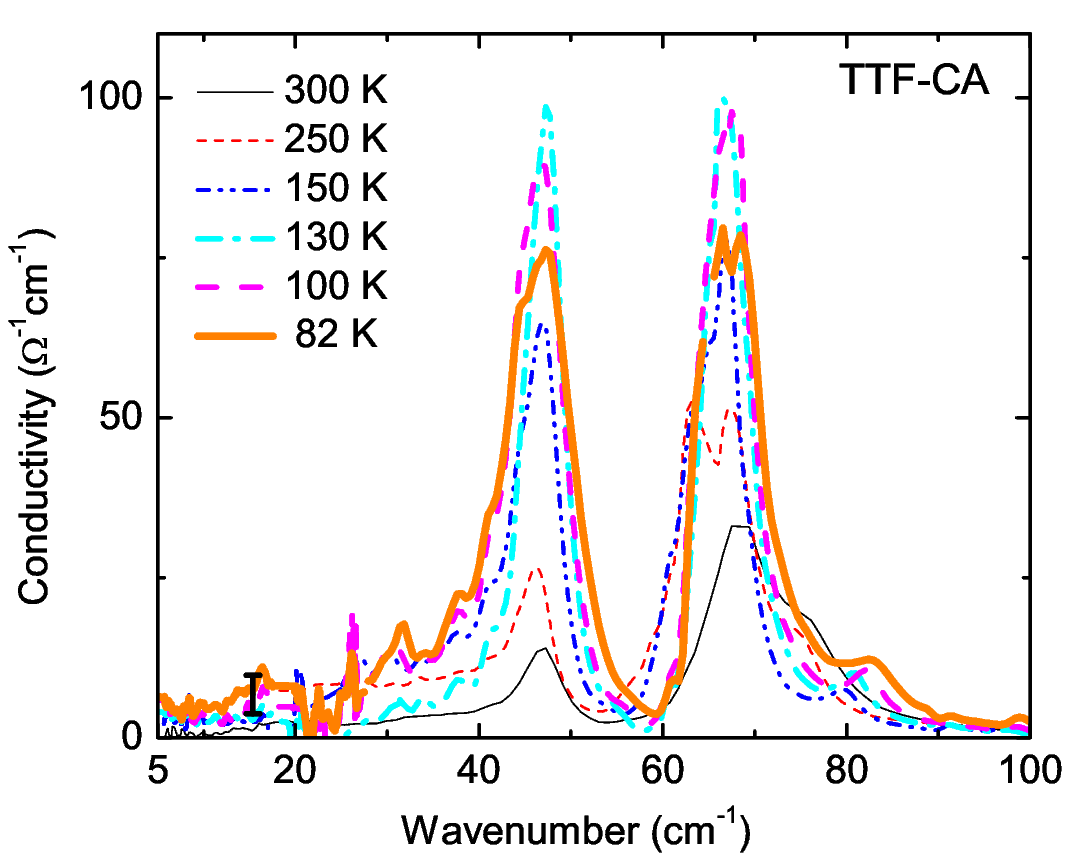}
\caption{(color online) Conductivity spectra along the $a$ stack
axis at temperatures above the {\it N -- I} transition. An error
bar for the absolute values of low frequency conductivity is
shown. }
 \label{fig:sigma_T}
\end{figure}

\begin{figure}
\includegraphics[width=8.5cm]{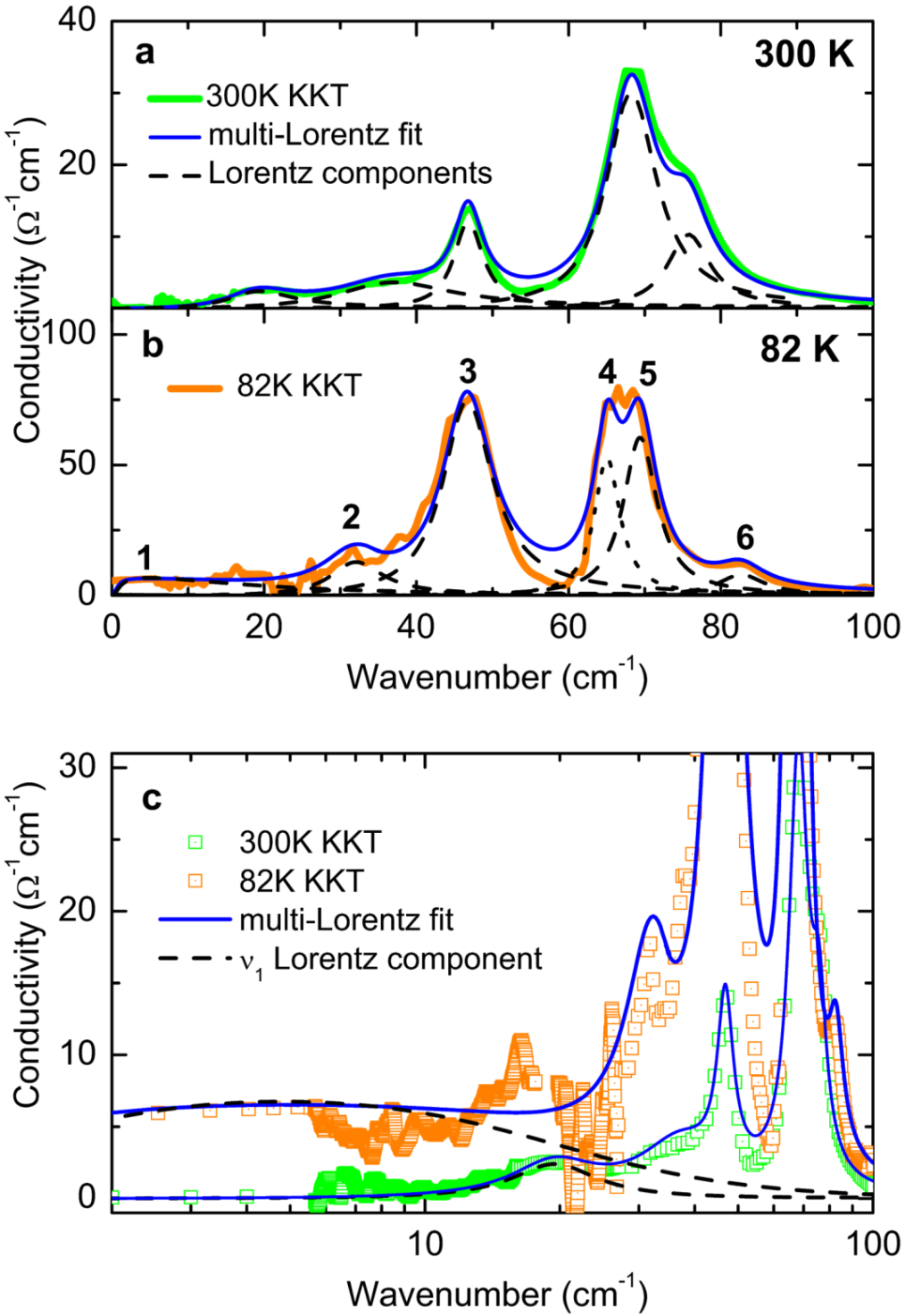}
\caption{(color online) Upper panel: Comparison of the
TTF-CA KKT conductivity at 300 K (green line) and 82 K (orange line)
with the fit of reflectivity spectrum by a minimum number of harmonic
oscillators. The Lorentzian bands used in the deconvolution are
shown as dashed lines, and the resulting conductivity as blue lines.
Bottom panel: Comparison of the KKT conductivity with the
multi-Lorentzian fit at two different temperatures. The
Lorentz component of the lowest frequency phonon ($\nu_1$)
is also shown. Notice the wavenumber logarithmic scale.} 
\label{fig:conductivity}
\end{figure}

Frequency dependent conductivity spectra, obtained by
the KKT of the reflectivity, are presented
in Fig.~\ref{fig:sigma_T}. In order to better understand the
temperature evolution of the conductivity spectra, and to gain
confidence in the band assignment, we have also fitted
the original reflectivity data with a set of Lorentzians.
The resulting conductivity practically coincides with the KKT
conductivity, as exemplified for the 300 and 82 K data in  
the upper panel of
Fig.~\ref{fig:conductivity}.
The individual Lorentzians used in the
deconvolution are also shown.
In the bottom panel of the Figure we report an enlarged view of the
temperature evolution of the KKT conductivity \textit{vs.}
the multi-Lorentzian reflectance fit. The logarithmic scale for the 
wavenumber axis focus attention on the deconvolution of the
lowest frequency phonon mode, that we shall discuss in detail below.
\begin{figure}
\includegraphics[width=8.5cm]{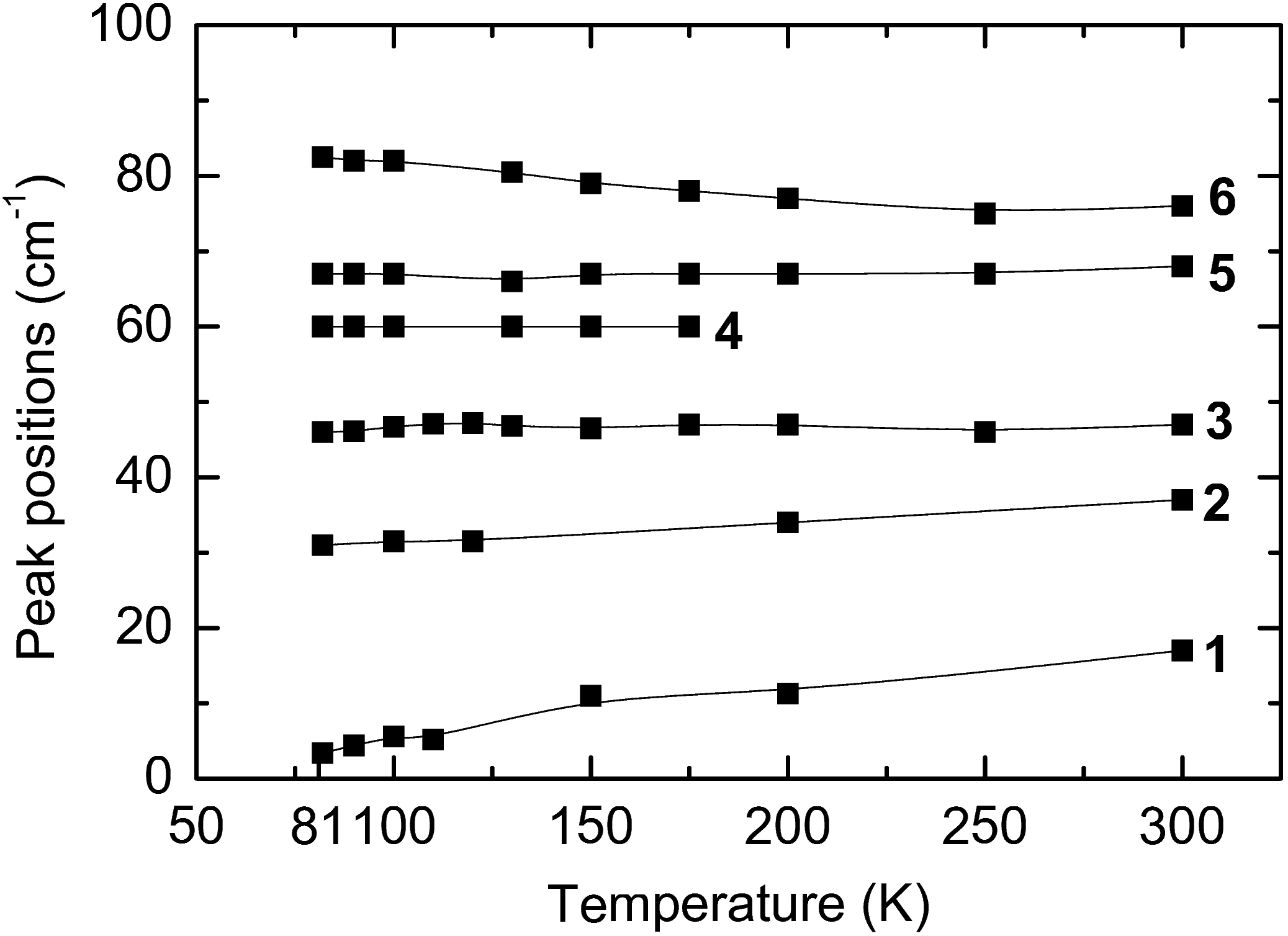}
\caption{Temperature dependence of the
frequencies of the six phonons bands observed below 100~\cm,
labeled $\nu_1$ to $\nu_6$ going from the lowest to the highest
frequency. }
\label{fig:exp_freqs}
\end{figure}

Six peaks are clearly identified in the spectral region below
100~\cm. We label them $\nu_1$ to $\nu_6$ in order of
increasing frequency, as shown in the upper panel of Fig.~\ref{fig:conductivity}.
The temperature evolution of the $\nu_1$ to $\nu_6$ peak frequencies
is reported in Fig.~\ref{fig:exp_freqs}.
In a group of bands with a broad maximum around 68~\cm\ at
$T=300$~K, we resolve three modes ($\nu_6, \nu_5$ and $\nu_4$) at
lower temperatures. While the high-frequency band shows normal
hardening on lowering $T$, the $\nu_5$ and $\nu_4$ modes
essentially do not change their positions. The $\nu_3$ band at
47~\cm\ exhibits a very weak softening of 2~\cm\ below $T=130$~K.
The deconvolution of the low frequency region of the room
temperature spectrum suggests the presence of bands
around 32 and 17~\cm\ ($\nu_2$
and $\nu_1$), ``appended'' to the 47~\cm\ band.
When going from ambient $T$ to 200~K, the $\nu_1$ band shifts
below 15~\cm\ and becomes wider. It shifts to even lower
frequencies and broadens as temperature is reduced, getting down
to about 5~\cm\ for $T=90$~K (Fig.~\ref{fig:exp_freqs}), 
assuming the shape of a broad background at temperatures just above the phase
transition (Fig.~\ref{fig:conductivity}, bottom panel).
\begin{figure}
\includegraphics[width=8.5cm]{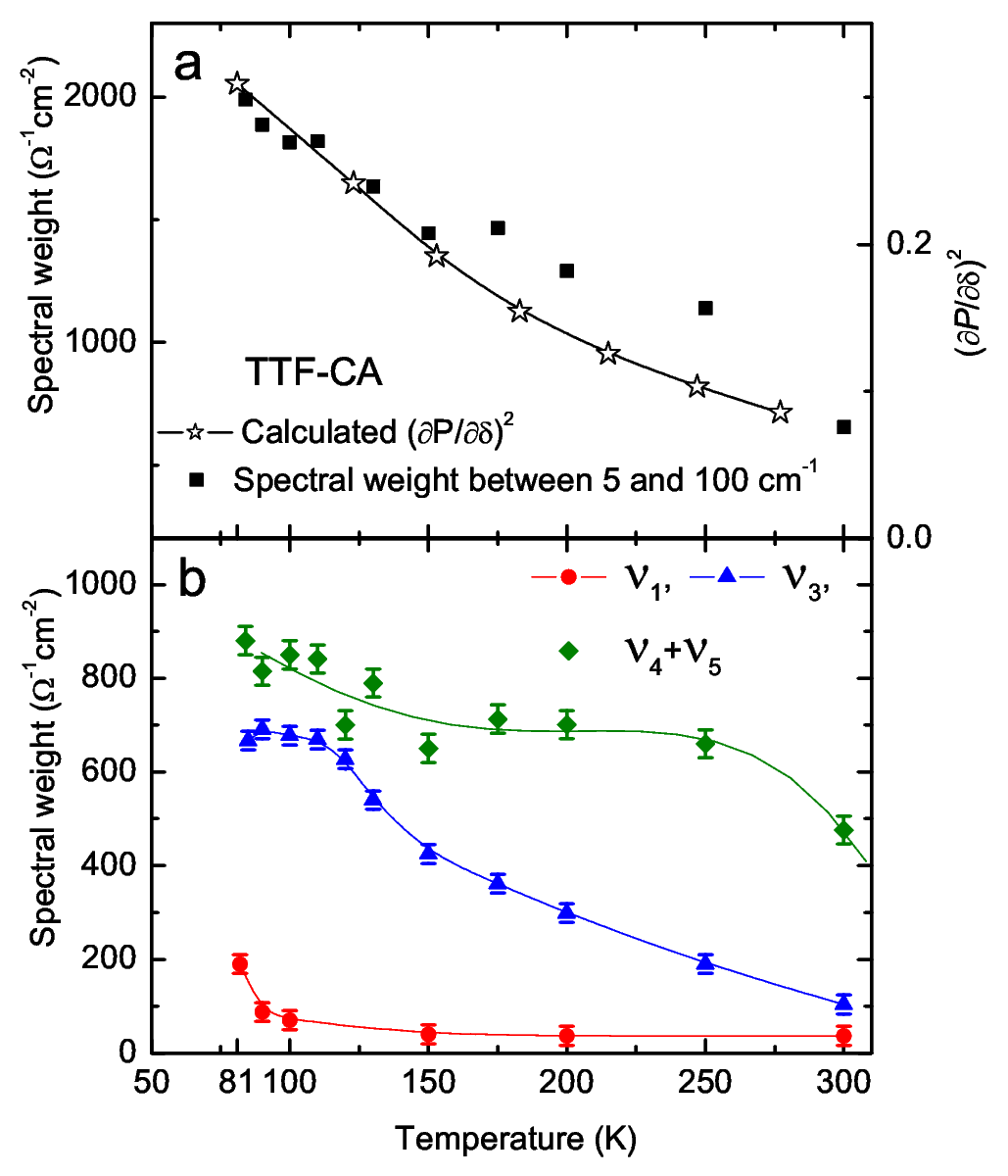}
\caption{(color online) (a) Temperature dependence of the spectral
weight of the $0-100$~\cm spectral region (squares), compared with
the temperature dependence of $(\partial P/\partial \delta)^2$
(open stars) (Section~\ref{sec:modelling}). (b) Temperature
dependence of the spectral weight of the group of bands around
70~\cm ($\nu_4$ and $\nu_5$) (rhombs), and of the $\nu_3$
(triangles) and $\nu_1$ (circles) bands. The lines are a guidance
to the eye.}
 \label{fig:exp_intens}
\end{figure}
\begin{figure}
\includegraphics[width=8.5cm]{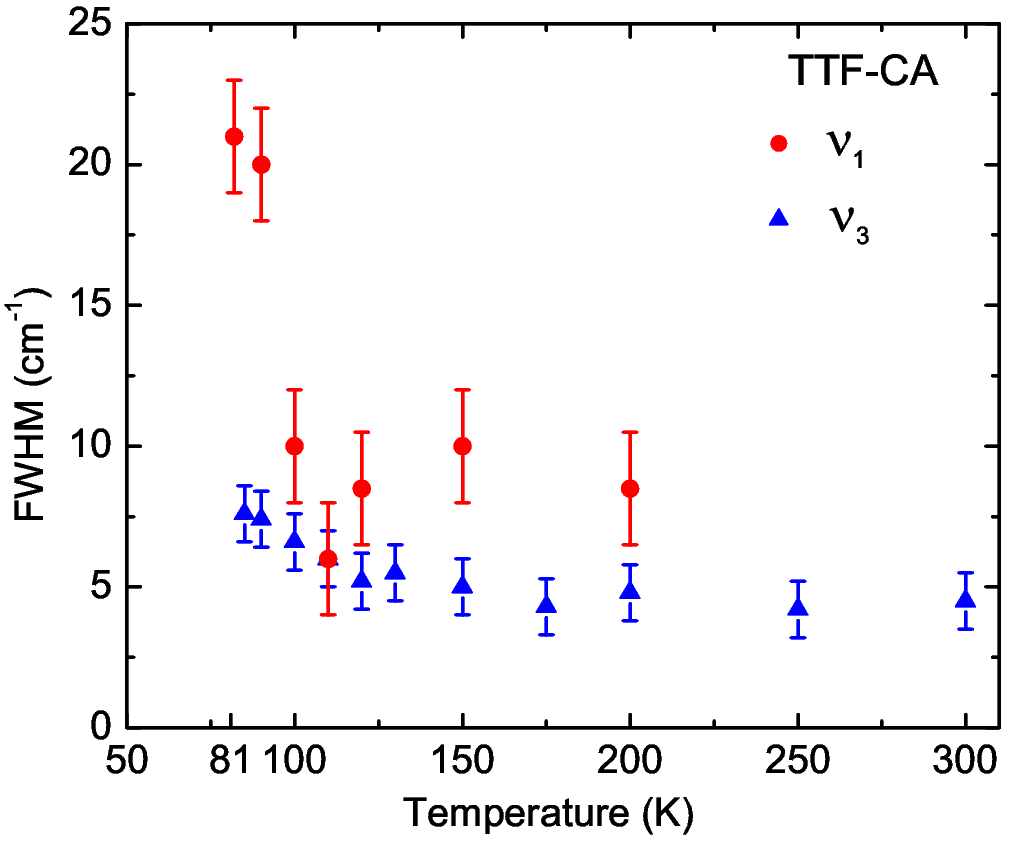}\\
\caption{(color online) Temperature dependence of the full width
at half maximum (FWHM) of the $\nu_1$ (circles) and of the $\nu_3$
(triangles) bands in TTF-CA.}
  \label{fig:halfwidth}
\end{figure}

The conductivity spectra in Fig.~\ref{fig:sigma_T} show a huge
increase of overall intensity by lowering $T$. A quantitative
estimate of this growth in terms of the spectral weight between 0
and 100~\cm\ is shown in Fig.~\ref{fig:exp_intens}a. The
intensities of the individual bands, estimated by integrating in
the relevant spectral region, are reported in the lower frame.
Fig.~\ref{fig:exp_intens}b shows that only some bands gain
intensity by lowering $T$: the $\nu_5$ and $\nu_4$ (estimated as a
whole maximum around 70~\cm\, since the bands are not well
resolved), the $\nu_3$ band at 47~\cm, and the $\nu_1$ band. The
intensity of the latter increases only below 100 K.

While the spectral weight increases on cooling, it also
redistributes towards lower frequency bands. The temperature
dependence of intensity around 70~\cm\ has smaller slope below
200~K, whereas the spectral weight of $\nu_3$ (47~\cm) band
saturates at temperatures below 130~K. At this temperature the
intensity of the $\nu_1$ band starts to rise rapidly. Just above
the transition, from 90 to 82~K, the $\nu_1$ band grows at the
expenses of the $\nu_3$ one.

With lowering temperature the $\nu_3$ and $\nu_1$ bands show
considerable anharmonicity. While at $T=300$~K the $\nu_3$ band
(47~\cm) can still be fitted with a Lorentzian, it becomes wider
and very asymmetric at temperatures below 250 K, as its lower
frequency wing grows on cooling. The temperature dependence of the
full width at half maximum (FWHM) of the $\nu_3$ band, directly
estimated from experimental spectra (Fig.~\ref{fig:conductivity}, bottom
panel), is displayed in Fig.~\ref{fig:halfwidth}. The same plot shows also the
dramatic increase of the FWHM of the lowest frequency $\nu_1$ band below $T$=100 K,
with clear \textit{overdamping} at temperatures close to the transition.

The present analysis of the spectra evidences a very specific
behavior of $\nu_5$, $\nu_4$, and especially $\nu_3$ and $\nu_1$
bands as the temperature approaches the NIT. The bands clearly
correspond to the most strongly coupled Peierls modes. By lowering
$T$, they show a huge intensity increase, with a concomitant
intensity redistribution, and important deviations from Lorentzian
bandshape. Below 100 K, we also observe appreciable softening and
large broadening for the lowest frequency mode. A quantitative
explanation of this complex temperature behavior of the spectrum
is offered by the model discussed in 
Section~\ref{sec:modelling}, while in the next Section we describe the
computational methods.

\section{Computational methods}
\label{sec:computational}

The separation of intramolecular vibrations from intermolecular,
or lattice, phonons is a common and useful approximation in
dealing with the complex phonon spectra of molecular crystals.
Thus lattice phonons describe translations and rotations of the
rigid molecules (rigid molecule approximation, RMA). In the framework
of a Hubbard-model description of the electronic structure (see
below), the electron-phonon coupling can also be separated into
two contributions: The totally symmetric molecular vibrations are
assumed to couple with electrons through modulation of on-site
energies (electron-molecular vibrational, or Holstein, coupling).
On the other hand, lattice phonons and possibly out-of-plane
molecular vibrations are expected to modulate the {\it
inter-}molecular CT integral (Peierls coupling).\cite{e-ph}

In order to characterize the Peierls coupling in TTF-CA, we have
first calculated the lattice phonon frequencies and normal
coordinates at $T=300$~K ({\it N} phase) by the quasi-harmonic
lattice dynamic (QHLD) method, as described in paper
I.\cite{paperI} The atom-atom potential adopted in QHLD accounts
for van der Waals and Coulomb forces, but does not include the CT
interaction. It should be noted that in the calculation we have
relaxed the RMA, by actually considering \textit{all}
the phonons below $\sim 250$~\cm.
\newpage

We describe the strength of Peierls coupling in terms of the linear
Peierls-coupling constants:
\begin{equation}
g_{i}=\sqrt{\frac{\hbar}{2\omega_i}} \left(\frac{\partial t_{\rm
DA}}{{\partial Q_{i}}}\right)_{eq} \label{eq:couplconst}
\end{equation}
\noindent where $t_{\rm DA}$ is the CT integral between adjacent
DA molecules along the stack, and $Q_{i}$ is the normal coordinate
for the $i$-th phonon with frequency $\omega_i$ and wavevector
$\textbf{q}=0$. The CT integral and its variation with $Q_{i}$ has
been calculated by the extended H\"uckel method, adopting the
Wolfsberg and Helmholtz approximation.\cite{EH} At the
room-temperature equilibrium geometry, we find $t$ = 0.20 eV, in
agreement with current estimates.\cite{soos04} The $g_{i}$ values,
obtained by numerical differentiation, offer a reliable indication
of the relative magnitude of the coupling constants. The Peierls
coupling strength of the $i$-th phonon is given by $\varepsilon_i
=(g_{i}^2/\omega_i)$, and the total coupling strength, or lattice
relaxation energy, is $\varepsilon_{\rm d} = \sum_i
\varepsilon_i$.

The electronic structure of TTF-CA is described in terms of a
modified Hubbard model with adiabatic coupling to molecular and
lattice vibrations. Real space diagonalization of the modified
Hubbard Hamiltonian relevant to stacks of up to 18 molecules with
periodic boundary conditions yields reliable information on the
ground state properties.\cite{soos04} The expectation value of the
dipole moment operator and its derivatives are calculated by the
Berry-phase approach.\cite{soos04}

\section{Spectral modeling}
\label{sec:modelling}

As discussed in Paper I \cite{paperI} and
elsewhere,\cite{moreac96,lecointe95} in the {\it N} phase of
TTF-CA the Peierls-coupled phonons transform as the $B_u$ species
in the $C_{2h}^5$ crystal symmetry, and are IR active with
polarization along the stack axis. We shall therefore restrict
our attention to these phonons. In Table~\ref{tab:Ncouplcon} we
list the QHLD calculated $B_u$ frequencies for the equilibrium
structure at 300~K,\cite{paperI} and the corresponding coupling
constants and coupling strength obtained as described in
Section~\ref{sec:computational}. The total coupling strength, or
lattice relaxation energy, is $\varepsilon_{\rm d} = 0.1$~eV.
Phonons above 250 \cm\ give negligible contribution to
$\varepsilon_{\rm d}$. As a consequence, they are not reported in
the Table nor discussed here.
\begin{table}[htp]
\caption{Calculated low-frequency $B_u$ phonons of TTF-CA in the
{\it N} phase. $\omega_i$, $g_i$, and $\varepsilon_i$ denote the
zero-order frequency, Peierls coupling constant, and coupling
strength, respectively.} \vskip 2mm
\begin{tabular*}{8cm}{@{\extracolsep{\fill}}lcrrr}
\hline \hline
&mode &$\omega_i$ (cm$^{-1}$) & ~$g_i$ (meV) & ${\varepsilon}_i$ (meV) \\
\hline
 $B_u$&$\nu_1$ & 27.6~~~ & $-5.16$~~~ & 7.8~~~ \\
      &$\nu_2$ & 38.6~~~ & $8.30$~~~ & 14.4~~~ \\
      &$\nu_3$ & 55.8~~~ & $-8.68$~~~ & 10.9~~~ \\
      &$\nu_4$ & 80.5~~~ & $-2.16$~~~ & 0.5~~~ \\
      &$\nu_5$ & 90.5~~~ & $-9.66$~~~ & 8.3~~~ \\
      &$\nu_6$ & 99.4~~~ & $-16.99$~~~ & 23.4~~~ \\
      &$\nu_7$ & 113.2~~~ & $-2.27$~~~ & 0.4~~~ \\
      &$\nu_8$ & 118.5~~~ & $-20.71$~~~ & 29.2~~~ \\
      &$\nu_9$ & 134.0~~~ & $0.08$~~~ & 0.0~~~ \\
      &$\nu_{10}$& 194.3~~~ & $7.85$~~~ & 2.6~~~ \\
      &$\nu_{11}$& 206.0~~~ & $-2.43$~~~ & 0.2~~~ \\
      &$\nu_{12}$& 211.3~~~ & $-2.93$~~~ & 0.3~~~ \\
      &$\nu_{13}$& 250.7~~~ & $-13.75$~~~ & 6.1~~~ \\
\hline \hline
\end{tabular*}
\label{tab:Ncouplcon}
\end{table}
As already mentioned in Section~\ref{sec:computational}, the QHLD
internuclear potential does not include the CT interaction, so
that values in Table \ref{tab:Ncouplcon} are the zero-order
reference frequencies in the absence of Peierls coupling. In other
words, they correspond to an hypothetical state with
$\varepsilon_d = 0$ or, equivalently, to a state where the
electronic excitations are moved to infinite energy.\cite{e-ph} As
a consequence, the frequencies of Table \ref{tab:Ncouplcon} cannot
be directly compared to the experimental frequencies. In fact,
phonons with high values of coupling constants are expected to be
the most intense in the spectra, and to occur at frequencies lower
than the zero-order ones. Therefore we can for instance anticipate
that the $\nu_6$ and $\nu_8$ modes of Table~\ref{tab:Ncouplcon}
correspond to the group of intense bands occurring around 70~\cm\
(Fig.~\ref{fig:sigma_T}). Incidentally, the eigenvectors of the
$\nu_6$ phonon correspond to rigid molecular displacements along
the stack, whereas the $\nu_8$ phonon has a more complex
description, being a mixture of TTF torsional motion and molecular
displacement along the stack. Modes above 200~\cm, on the other
hand, are almost pure intramolecular modes, like for instance the
$\nu_{13}$ phonon, which essentially corresponds to a TTF
out-of-plane motion. Their intensity may then have a significant
``intrinsic'' contribution (namely, not due to Peierls coupling),
that is not accounted for in the present model calculation.

Both experiment and calculations suggest that several modes are
appreciably coupled to the CT integral, leading to a
\textit{multi-mode} Peierls coupling. The resulting problem
becomes complex because the phonons are mixed through their common
interaction with the CT electrons. Thus a meaningful
comparison with experiment requires detailed modeling. The
effect of Peierls coupling can be dealt with separately for the
phonon frequencies and for their IR intensity.\cite{soos04}
Accordingly, the analysis will be carried out in two steps.
\begin{table}
 \centering
 \caption{Calculated electronic response $\chi_b$
and squared polarizability derivative $(\partial P/\partial
\delta)$ for different degree of ionicity $\varrho$. The
correspondence between $\varrho$ and $T$ is derived from
experiment (Ref.~\onlinecite{sidebands}).}
\begin{tabular*}{8cm}{@{\extracolsep{\fill}}llccr}
\hline \hline
~$\chi_b$ (eV$^{-1}$) & $(\partial P/\partial \delta)^2$ & $\varrho$~~~&& $T$ (K) \\
\hline
 ~ 5.7971 & ~ 0.0858 & 0.197 && 277 \\
 ~ 6.1230 & ~ 0.1034 & 0.209 && 247 \\
 ~ 6.4862 & ~ 0.1257 & 0.222 && 215 \\
 ~ 6.8944 & ~ 0.1544 & 0.236 && 183 \\
 ~ 7.3561 & ~ 0.1919 & 0.250 && 153 \\
 ~ 7.8837 & ~ 0.2415 & 0.266 && 123 \\
 ~ 8.4927 & ~ 0.3089 & 0.282 && 81 \\
\hline \hline
\end{tabular*}
\label{tab:dvb}
\end{table}

It has been shown that in the presence of Peierls electron-phonon
interaction, the squared perturbed frequencies $\Omega_j^2$ and
normal modes coordinates $\mathcal{Q}_j$ are obtained from the diagonalization
of the following force constant matrix, written in the basis of the
reference coordinates $Q_i$:\cite{pg88}
\begin{equation}
F_{ij}=\omega_{i} \omega_{j} \delta_{ij}-\sqrt{\omega_{i}
\omega_{j}}~ g_i g_j~\chi_b .
\label{eq:fmatrix}
\end{equation}
In this equation, $\omega_i$ and $g_i$ are the reference frequencies and
coupling constants of Table \ref{tab:Ncouplcon}, and $\chi_b$ is
the electronic response to phonon perturbation. The $\chi_b$
values needed to apply Eq.~\ref{eq:fmatrix} are calculated as
described in Ref. \onlinecite{soos04} by tuning the parameters of
the modified Hubbard model as to mimic the behavior of TTF-CA.
Since in going from 200 to 82 K the TTF-CA ionicity $\varrho$
changes from about 0.2 to 0.3;\cite{girlando83,sidebands} in the
first column of Table \ref{tab:dvb} we list the $\chi_b$ values
corresponding to different ionicities in this interval. By
matching the calculated ionicities in the third column of the
Table with the experimental ones,\cite{sidebands} we can translate
the $\varrho$-dependence of the calculated $\chi_b$ in a
$T$-dependence, as reported in the fourth column of Table
\ref{tab:dvb}. In this way we can perform a more direct comparison
with experimental data.

\begin{figure}
 \includegraphics[width=8.cm]{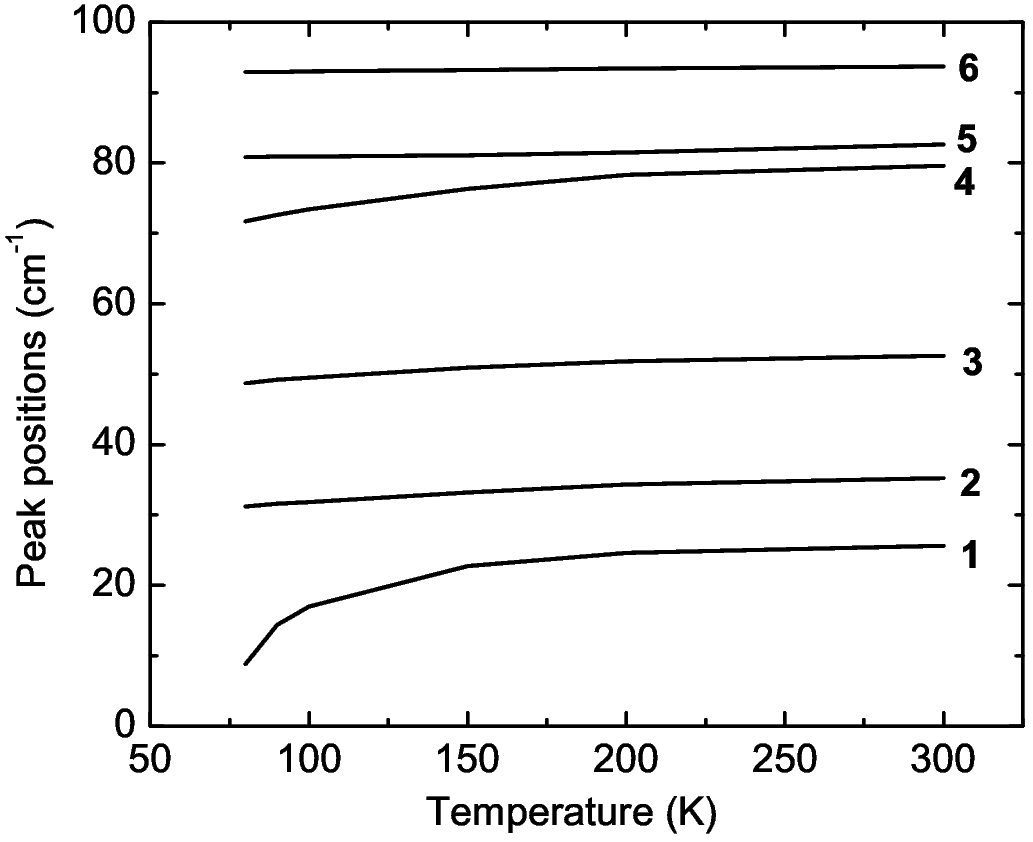}
\caption{Temperature dependence of the Peierls-coupled modes,
calculated by Eq.~(\ref{eq:fmatrix}), on the basis of
Table~\ref{tab:Ncouplcon} and Table~\ref{tab:dvb} parameters.}
  \label{fig:multiPeierls}
\end{figure}

Fig.~\ref{fig:multiPeierls} summarizes the temperature dependence
of the perturbed frequencies $\Omega_j$ below 100~\cm\, calculated
through Eq.~(\ref{eq:fmatrix}), with the parameters listed in
Tables~\ref{tab:Ncouplcon} and \ref{tab:dvb}. The agreement 
between the calculated frequencies in Fig.~\ref{fig:multiPeierls} and
the experimental ones presented in Fig.~\ref{fig:exp_freqs} is very good.
First of all, the number of modes below 100~\cm\ is the same, namely six.
Second, an appreciable softening is detected only for the lowest
frequency mode, and in the proximity of the phase transition. The
weak softening shown by some higher frequency modes in
Fig.~\ref{fig:multiPeierls} is likely compensated by the usual
frequency hardening by lowering $T$, a factor not included in the
model.

The calculations quite naturally explain why -- although several
phonons are coupled to the CT electrons -- the red shift of
the associated bands on getting close to the transition is so
small. As mentioned above, all the phonons are coupled together
through their common interaction with the electronic system. Then,
when two phonon frequencies get closer due to a softening of the
higher frequency phonon, there is mixing in the phonon
description, and the softening is ``transferred'' to the lower
frequency phonon. In more precise terms, with lowering temperature
$\chi_b$ increases with $\varrho$ (Table~\ref{tab:dvb}), leading
to an evolution of the normal modes $ \mathcal{Q}_j$:
\begin{equation}
\mathcal{Q}_j = \sum_i l_{ij} Q_i ,
\label{eq:eigenvectors}
\end{equation}
where $l_{ij}$ are the eigenvectors obtained, for each
temperature, by diagonalizing the \textbf{\textit{F}} matrix of
Eq. \ref{eq:fmatrix}. The Peierls coupling constants are
accordingly modified: The coupling constants $G_j$ in the basis of
the perturbed normal modes $\mathcal{Q}_j$ are linear combination
of the reference coupling constants:
\begin{equation}
G_j = \sum_i l_{ij} g_i \sqrt{\frac{\omega_i}{\hbar}} ,
\label{eq:COUPLCON}
\end{equation}
and therefore evolve with $\chi_b$. As a result, on approaching
the phase transition the coupling
strength is progressively transferred to the lower frequency modes.
Only in close proximity of
the transition all the coupling strength collapses to the lowest
frequency phonon, yielding substantial softening of this specific
mode.

We now turn our attention to the intensity of the phonon bands.
For a regular stack (equispaced D and A units along the stack,
like \textit{N} TTF-CA) the IR intensity of the Peierls modes is
largely dominated by a term accounting for the charge fluctuations
induced by oscillation in the dimerization amplitude $\delta$. For
a single Peierls mode with frequency $\omega_\mathrm{P}$ the total
oscillator strength is:\cite{soos04}
\begin{equation}
f_{\mathrm{P}} = \frac{m_e d^2 \omega_{\mathrm{P}}^2
\varepsilon_{\rm d}}{t^2} \left(\frac{\partial P}{\partial
\delta}\right)^2 , \label{eq:zolt_anna}
\end{equation}
where $m_e$ is the electronic mass, $d$ the equilibrium distance
between D and A molecules, and $P$ the electronic polarizability
per site. The $(\partial P / \partial \delta)^2$ values calculated
for parameters relevant to \textit{N} TTF-CA are reported in the
second column of Table~\ref{tab:dvb}. The temperature dependence
of the $(\partial P/\partial \delta)^2$ terms compares well with
the experimentally determined growth of the spectral weight in the
range of 0-100 \cm\ (Fig.~\ref{fig:exp_intens}).

In the multi-mode Peierls coupling case, the oscillator strength
is partitioned among the coupled modes. Specifically, each normal
coordinate $\mathcal{Q}_j$ modulates the CT integral as described
by the coupling constants $G_j$ in Eq.~(\ref{eq:COUPLCON}). Using
the usual chain-rule, the $\delta$-derivative of $P$ in
Eq.~(\ref{eq:zolt_anna}) can be rewritten as a sum of
$\mathcal{Q}_j$ derivatives. After some algebra we derive the
following expression for the IR oscillator strength of each mode
coupled to the electronic degrees of freedom:
\begin{equation}
 f_j = \frac{m_e d^2}{t^2}~ G_j^2
 \left(\frac{\partial P}{\partial \delta}\right)^2 .
 \label{eq:oscillator}
\end{equation}
We can now use the standard expression for the frequency-dependent
dielectric constant
\begin{equation}
\epsilon (\omega) = \epsilon_\infty + 8.96857\cdot 10^{10}~
\frac{N}{V}\sum_k \frac{f_j}{(\Omega_j^2 -
\omega^2)-\mathrm{i}\omega \Gamma_j} \label{eq:epsilon}
\end{equation}
to calculate the contribution of the Peierls coupled modes to the
low-frequency spectra of TTF-CA. The numerical factor in
Eq.~(\ref{eq:epsilon}) refers to $N/V$ (number of molecules per
unit-cell volume) expressed in \AA$^{-3}$, and the frequencies
$\Omega_j$, $\omega$, and damping $\Gamma_j$, in \cm.

The calculated reflectivity is compared to the experimental one in
Fig.~\ref{fig:allRefl_300K} of Section \ref{subsec:reflspectra}.
The unperturbed frequencies and coupling constants needed for the
calculation are taken from Tab.~\ref{tab:Ncouplcon}, whereas
$\Gamma_j$ is set equal to 4.0 \cm~for all the modes. The values
of $\chi_b$ = 5.6667 eV$^{-1}$ and $(\partial P / \partial
\delta)^2$ = 0.07922 are obtained from interpolation of the data
in Tab.~\ref{tab:dvb}. In addition, we use $d = 3.7$~\AA, $N/V =
4.92386 \cdot 10^3$~\AA$^{-3}$ (Ref.~\onlinecite{lecointe95}), and
$t = 0.2$~eV. The CT electronic transition is added in the
calculation as an extra term in Eq.~(\ref{eq:epsilon}), where
$f_{\rm CT}$, $\Omega_{\rm CT}$ and $\Gamma_{\rm CT}$ are derived
directly from experiment.\cite{jacobsen83} Finally, we put
$\epsilon_\infty = 2.5$ to account for the high-energy
contributions.

Fig.~\ref{fig:allRefl_300K} shows a very good agreement between
model calculation and experiment, considering all the
approximations involved in the model and in the estimate of the
parameters, in particular of the $g_i$. The frequencies are
somewhat off, but well within the usual errors of QHLD
calculations ($\pm$ 10 \cm). The calculated oscillators strengths
for the lower frequency modes are just a little below the
experiment. The agreement for the oscillator strength of the
phonons at higher frequencies is not expected to be exact,
because these phonons have intra-molecular character and, as
discussed above, may have a non-negligible ``intrinsic''
intensity. The simulation indicates that at room temperature the
most strongly coupled modes cluster around approximately
$70-80$~\cm.
\begin{figure}
 \centering
  \includegraphics[width=8.5cm]{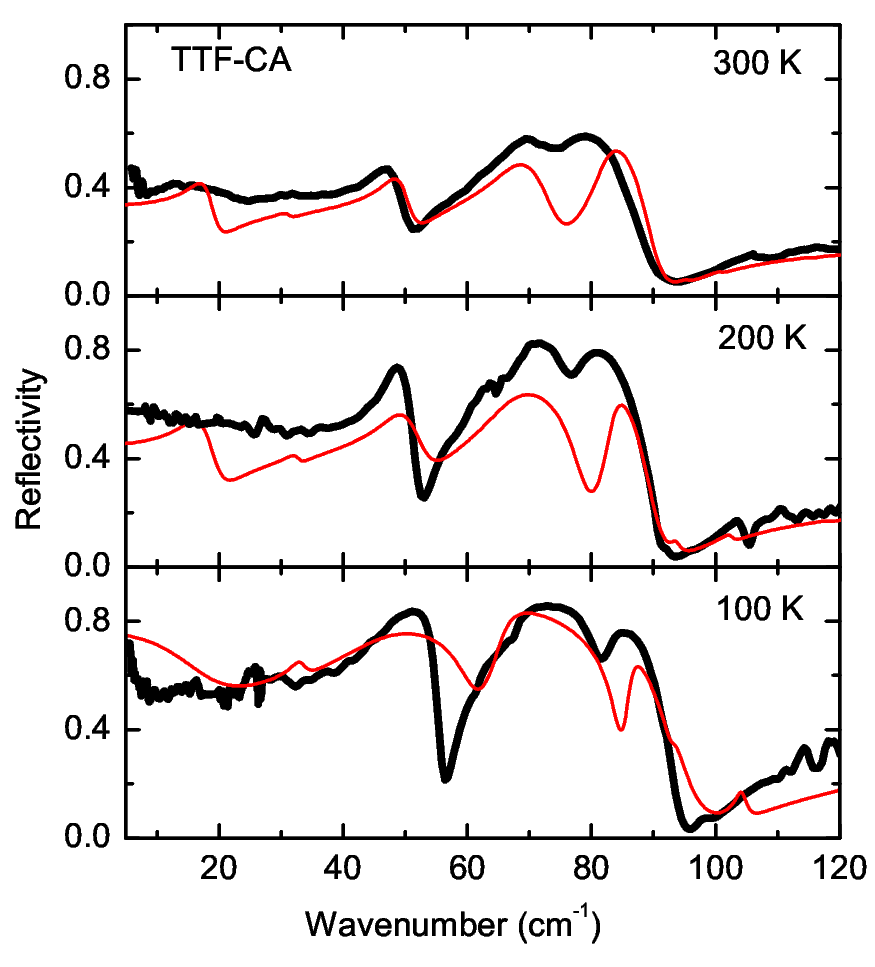}
 \caption{Experimental (thick line) and calculated (thin line) low-frequency reflectivity
spectra of TTF-CA, parallel polarization, at three different
temperatures.}
 \label{fig:calexp_spectra_T}
\end{figure}
\begin{figure}
 \centering
 \includegraphics[width=8.5cm]{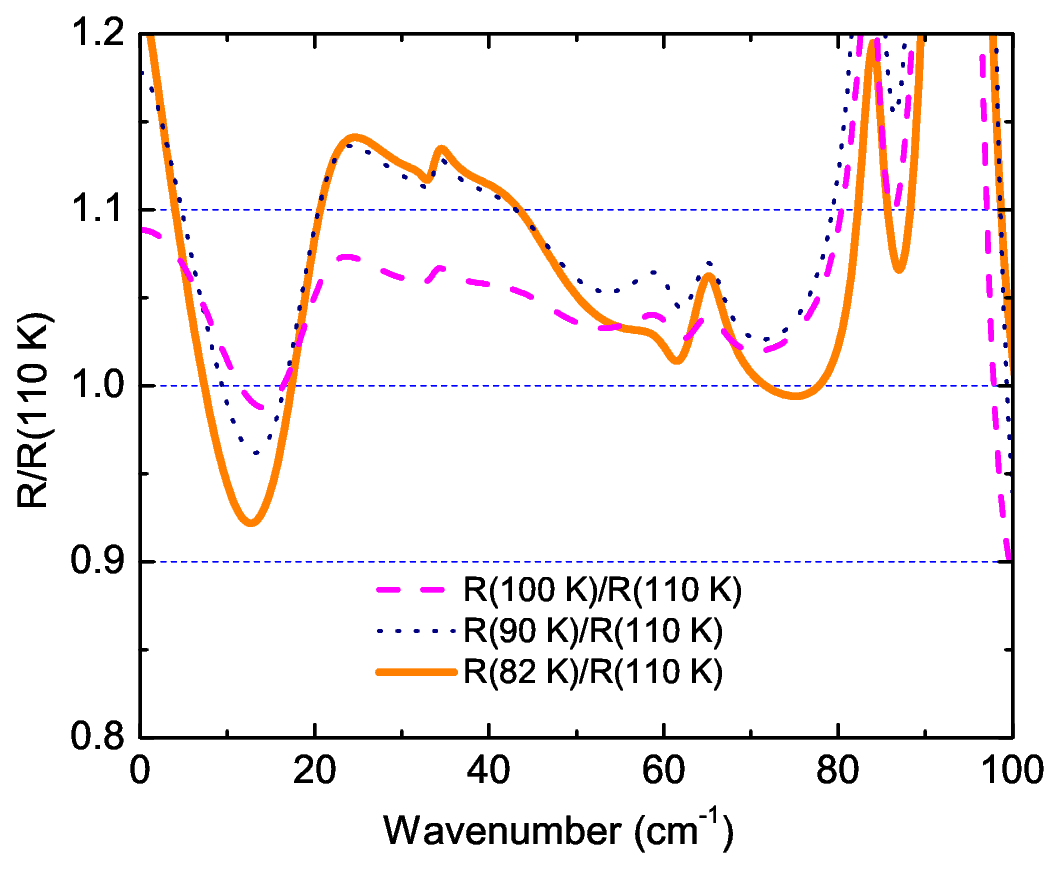}
 \caption{(color online) Calculated ratio of reflectance of the spectra below 110 K
to the spectrum at 110 K.}
 \label{fig:divide_calc}
\end{figure}

The discussion in Section \ref{sec:expspectra} puts in evidence a
quite complex evolution of TTF-CA spectra with temperature. Such
behavior is qualitatively reproduced by the calculation. An
overall increase of reflectivity by lowering $T$ is brought in by
the increase in $(\partial P / \partial \delta)^2$, and the shift
of spectral weight towards lower frequency is accounted for by the
change in the coupling constants $G_j$, as discussed after
Eq.~(\ref{eq:COUPLCON}). However, we can gain a better insight
into the factors affecting the temperature evolution of the
spectra if better agreement between experiment and simulation is
achieved. To such aim, Tab.~\ref{tab:Ncouplcon} frequencies and
coupling constants are used as starting values for a nonlinear
fitting procedure based on Eqs.~(\ref{eq:fmatrix}) and
(\ref{eq:COUPLCON}) to (\ref{eq:epsilon}).

The variation of the frequencies and of the oscillator strengths
with increasing $\varrho$ (lowering $T$) is induced in the model
by the corresponding increase in both $\chi_b$ and $(\partial P /
\partial \delta)^2$, as detailed in Table~\ref{tab:dvb}. However,
it is clear from experiment that also the bandwidths (damping)
$\Gamma_j$ change with $T$. In the model, the bandwidths are
additional parameters, but in the fit we choose to correlate them
to the anharmonicity introduced by the electron-phonon coupling.
Following Fano,\cite{fano} we impose the following relationship to
$\Gamma_j$ (in \cm):
\begin{equation}
\Gamma_j = \mathtt{k}~~ \frac{G_j^2}{\Omega_j} + 2.0 = \mathtt{k}~
\eta_j + 2.0. \label{eq:gamma}
\end{equation}
On top of the intrinsic linewidth of 2.0~\cm, the electron-phonon
coupling gives an additional contribution proportional to $\eta_j
= G_j^2/\Omega_j$. The proportionality factor $\mathtt{k}$ adds
just one extra adjustable parameter to the fit, and
Eq.~(\ref{eq:gamma}) represents a strong constrain. The fit shows
a high sensitivity of spectra on the adjustable parameters. Thus a
variation of the unperturbed frequencies $\omega_i$ or of the
coupling constants $g_i$ by a few per cent, may profoundly alter
the shape of the calculated spectra, especially in the case of the
strongly coupled modes which cluster around $70-80$~\cm. For this
reason, it is a challenge to obtain a satisfactory fit at all
temperatures.

Fig.~\ref{fig:calexp_spectra_T} illustrates the results of the fit
of the reflectivity spectra at three temperatures. The fit is
restricted to the spectral region below 150~\cm, so that the CT
transition and its variation with temperature\cite{jacobsen83} is
simulated by $\epsilon_\infty$. Furthermore, we allow the
frequencies $\omega_i$ to increase by a few wavenumber on lowering
$T$, in order to simulate the usual hardening due to thermal
contraction. The coupling constants and the proportionality factor
for the damping, $\mathtt{k}$, are instead kept temperature
independent.

Albeit not all details of the spectrum are precisely reproduced,
the essential features are caught: By lowering the temperature the
reflectivity increases, with a redistribution of the oscillator
strengths and of the bandwidths as the spectral weight shifts
towards lower frequencies. The downshift of spectral weight
becomes more dramatic just before the phase transition. This is
clearly demonstrated in Fig.~\ref{fig:divide_calc}, which
resembles the experimental behavior reported in
Fig.~\ref{fig:dividing}. The shift of the spectral weight is
accompanied by an increase of the bandwidth of the $\nu_1$, lowest
frequency phonon, which becomes overdamped around 100 K and below
(Fig.~\ref{fig:calexp_spectra_T}). The experimentally observed
overdamping of the $\nu_1$ phonon in Fig.~\ref{fig:halfwidth} is
then explained by the transfer of the entire electron-phonon
coupling strength to this phonon just before the phase transition,
with associated increase of anharmonicity.

\section{Discussion and conclusions}
\label{sec:concl}

In this paper we have presented temperature dependent polarized
reflectivity spectra collected down to 5~\cm\ for TTF-CA crystals
above the NIT. A detailed theoretical analysis of the
data has allowed us to offer the first direct identification
of the Peierls softening mechanism.

\begin{figure}
 \centering
 \includegraphics[width=8.0cm]{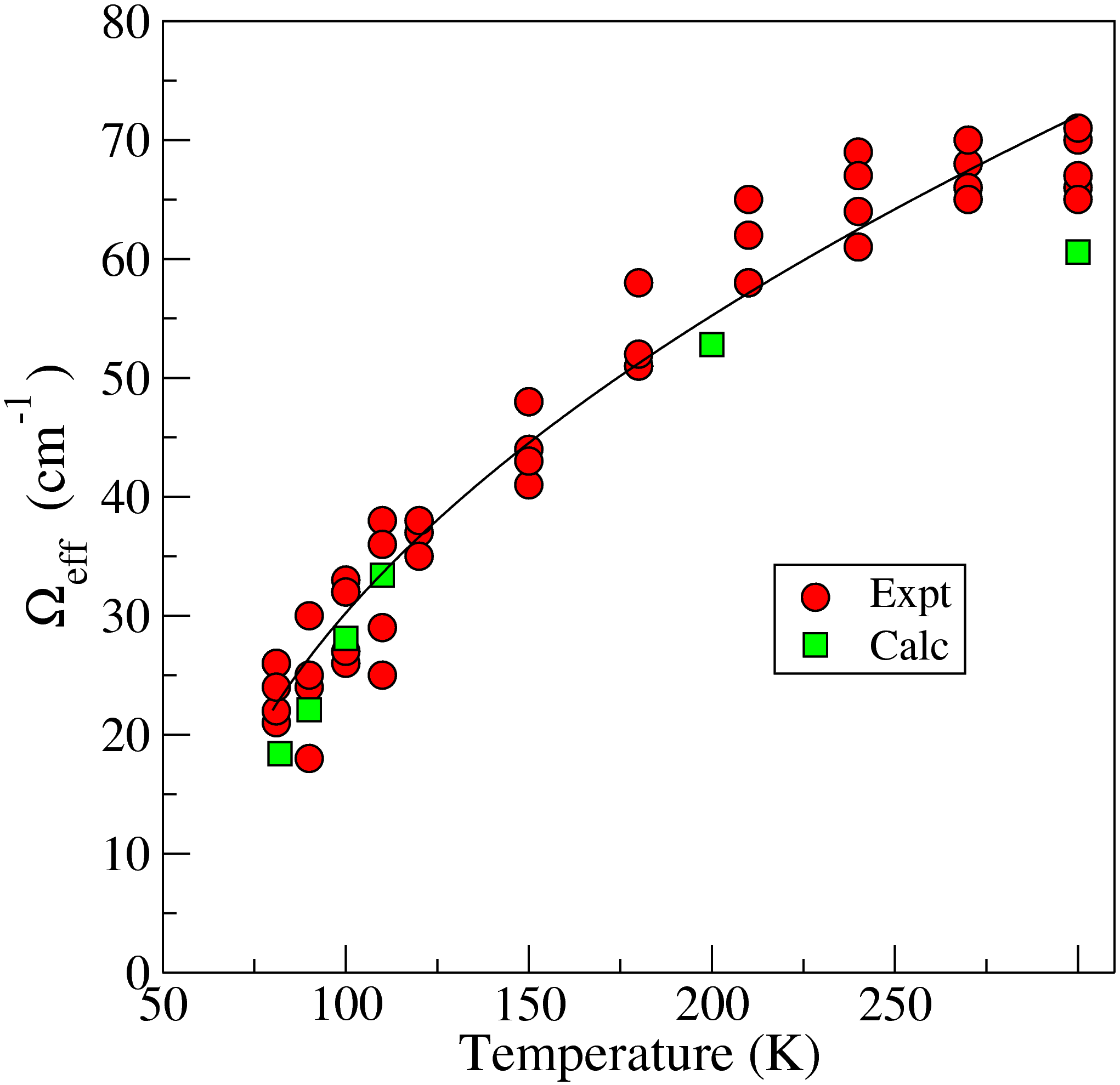}
 \caption{(color online) Experimental (taken from Ref.~\onlinecite {sidebands})
and calculated frequency (squares) of the effective soft mode.}
 \label{fig:effective_Peierls}
\end{figure}

An indirect experimental evidence of the softening of lattice
vibrations in TTF-CA was offered by the occurrence of combination
(two-phonon) bands in vibrational spectra collected in the region
of the intramolecular vibrations (mid-IR).\cite{sidebands} Several
broad bands were found in IR spectra of TTF-CA that could
not be assigned to fundamental modes. These bands, whose intensity
increases on approaching NIT, are symmetrically located at the low
and high frequency side of a Raman band assigned to a
totally-symmetric molecular vibration. The two side-bands are then
assigned to the sum and difference combinations of the
totally-symmetric molecular vibration with a lattice phonon. The
frequency of this lattice phonon lowers from about 70~\cm\ to
about 20~\cm\ when going from room temperature to 81~K, as
illustrated by the dots in fig. \ref{fig:effective_Peierls}. In
the light of the present work, indicating a multi-mode Peierls
coupling, the soft-phonon inferred from the analysis of
combination bands must correspond to an ``effective'' Peierls
phonon, resulting from the weighted contribution of the several
Peierls-coupled modes.

Specifically, the side-bands in the mid-IR region result from the
combination of a totally symmetric molecular vibration with a
superposition of several $B_u$ modes. Thus the apparent peak
frequency of the effective soft mode, $\Omega_{\rm eff}$, is the
weighted average of the frequencies of the Peierls phonons. In
order to calculate $\Omega_{\rm eff}$ from the spectral simulation
discussed in the previous Section, we adopt the following
expression:
\begin{equation}
\Omega_{\rm eff} = \frac{\sum_j \eta_j \Omega_j}{\sum_j \eta_j} ,
\label{eq:effective}
\end{equation}
where each frequency is weighted by $\eta_j$, the same factor
entering the definition of vibrational linewidth in
Eq.~(\ref{eq:gamma}), and connected to the strength of the Peierls
coupling. The computed value of $\Omega_{\rm eff}$ is compared
with experiment in Fig.~\ref{fig:effective_Peierls}. The very good
agreement between the effective phonon frequency obtained from the
analysis of combination bands in the mid-IR region and of lattice
phonons in the far-IR region strengthen our interpretation of the
spectroscopic data in terms of a soft mode. Quite interestingly,
the same effective soft mode of TTF-CA quantitatively accounts for
the peaks observed in the diffuse X-ray scattering
data.\cite{davino07}

\begin{figure}
\centering
\includegraphics[width=8.0cm]{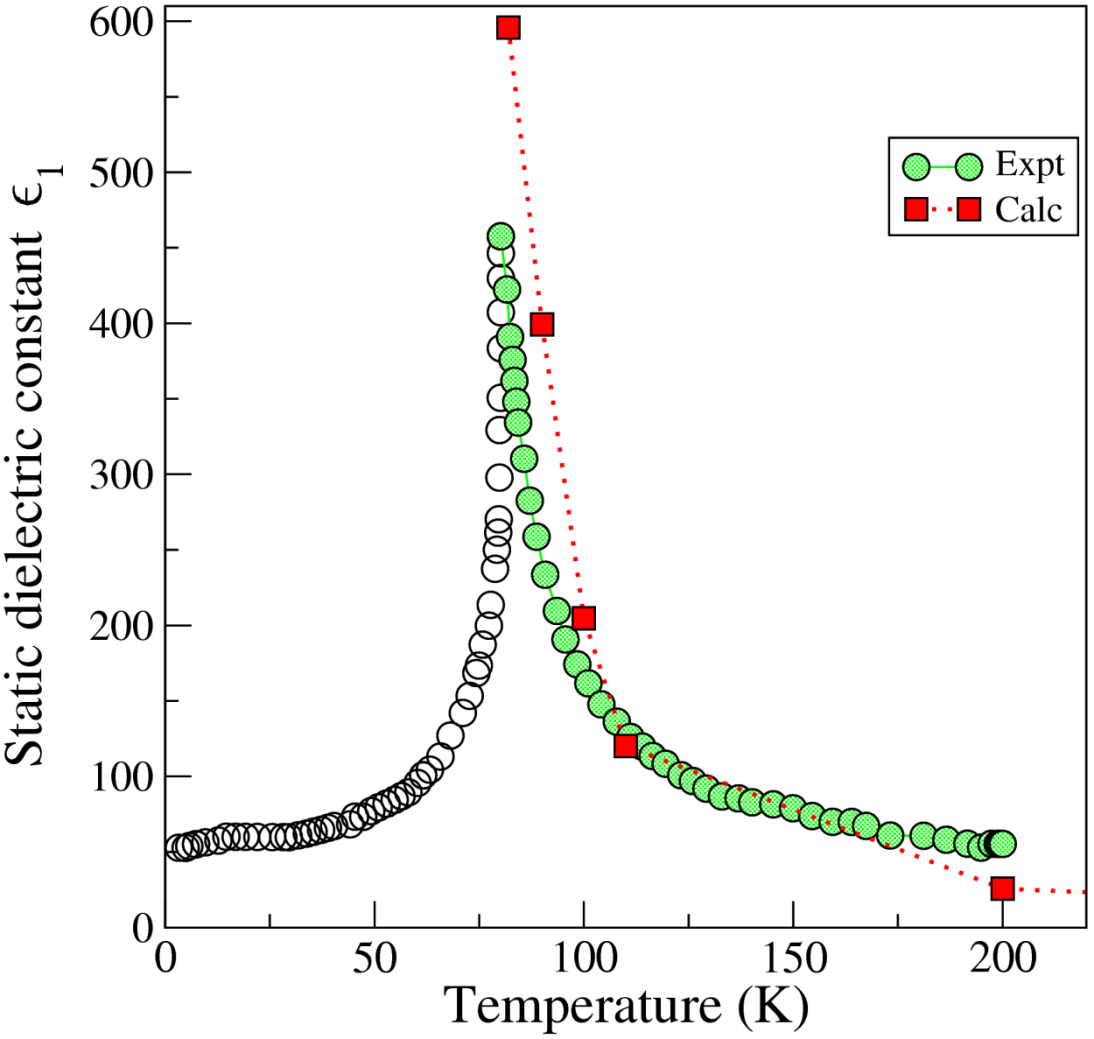}
\caption{(color online) Experimental (taken from Ref. \onlinecite{TTFCAdiel})
and calculated [Eq.~(\ref{eq:epsilon})] static dielectric constant of TTF-CA as
a function of temperature.}
 \label{fig:dielectric}
\end{figure}

The softening of the lattice phonons, combined with the increase
of their intensity on approaching NIT leads to very large
vibrational contributions to the dielectric
constant.\cite{freo02,soos04} By using the parameters found in the
fit of reflectivity spectra, we can also estimate the temperature
evolution of the static dielectric constant, $\epsilon_1$, which
corresponds to the zero-frequency real part of the dielectric
constant in Eq.~\ref{eq:epsilon}. As shown in
Fig.~\ref{fig:dielectric}, the present estimate of $\epsilon_1$
agrees very well with available experimental data, strongly
supporting the vibrational origin of the dielectric anomaly at
NIT, as due to the large charge oscillations associated with the
Peierls modes.\cite{freo02,soos04}

Data analogous to the present ones have been previously reported
Okimoto \textit{et al.},\cite{okimoto01} who discussed the
reflectivity spectra along the stack of TTF-QBrCl$_3$ (QBrCl$_3$:
2-bromo-3,5,6-trichloro-$p$-benzoquinone), a CT crystal similar to
TTF-CA that undergoes the NIT with continuous evolution of
ionicity. However, in that work the reflectivity data have been
collected only from 650 down to 25~\cm, so that the subsequent
KKT transformation is delicate, and the discussion of
conductivity spectra outside the experimentally accessed region
should be taken with caution. In any case a broad band was
enucleated from conductivity spectra which softens, without major
broadening, from about 60 \cm\ at 293 K to about 10 \cm\ just
above the critical temperature (71 K). This band was not ascribed
to a soft mode, but rather to the pinned mode of the so-called
neutral-ionic domain walls (NIDW).\cite{okimoto01} Whereas more
extensive measurements are required to fully address this issue,
our work on TTF-CA sheds doubts on this interpretation.

NIDWs were introduced theoretically by Nagaosa \cite{nagaosa86} as
charged boundaries of \textit{I} dimerized domains excited in the
host \textit{N} regular chain. Quite interestingly, they were
originally restricted to the close proximity of discontinuous NIT,
while the energy of the domains is often too large to allow for
thermal population.\cite{soos07} However, NIDWs have been quite
often invoked to explain the rich phenomenology of
NIT.\cite{ricemem,tokurareview} In particular, the above mentioned
dielectric anomaly and the combination bands in mid-IR spectra
were initially attributed to NIDWs,\cite{IR-NIDW} as
well as the peak in the diffuse X-ray signal.\cite{buron06} As
discussed above, this paper gives definitive support to a
different picture that, without invoking exotic excitations,
quantitatively explains the rich and variegated phenomenology of
NIT to the increase of the effective Peierls coupling on
approaching NIT. In this way different and apparently unrelated
phenomena are all naturally and \textit{quantitatively} explained
in terms of the softening of lattice phonons coupled via a Peierls
mechanism to delocalized electrons.

The definitive and unambiguous support to the above picture is
obtained by a careful analysis of the far-IR spectral region to
directly identify the Peierls modes. The analysis is non-trivial:
apart from the requirement for very refined experimental data
extending down to very low-frequency, the spectral interpretation
is based on a model for multi-mode Peierls coupling that combines
lattice dynamics (QHLD) calculations with the modified
Hubbard-model for the electronic structure. In fact, in the presence of
several phonons the multi-mode Peierls softening is a
complex phenomenon: the mode description changes on approaching
the transition, with associated mode mixing, intensity redistribution and
broadening. In the course of this mixing, the softening ``jumps''
from one mode to the nearest one below, until in the proximity of
the phase transition all the softening is transferred to the
lowest frequency mode. The mechanism, evident in the simulation
presented in Fig.~\ref{fig:multiPeierls}, explains the
softening observed in Fig.~\ref{fig:exp_freqs} and the temperature
behavior of mode intensities and linewidths in
Fig.~\ref{fig:exp_intens}. The increase of the electronic
susceptibility as the system is driven towards the NIT implies an
increased effective coupling, and accordingly the spectral weight
shifts towards zero frequency for $T<100$~K. Indeed, in this
temperature range the Peierls coupling is almost completely
transferred to the lowest-frequency mode. At the same time, the
relevant bandwidth broadens, finally leading to an overdamped
behavior. For this reason, in the spectra we just see the
reflectivity increase towards zero frequency
(Figs.~\ref{fig:reflect_T} and \ref{fig:dividing}), without being
able to actually detect the full absorption. Overdamping has been
often observed in ferroelectric phase
transitions.\cite{nakamura66} In the case of TTF-CA, we are able
to provide the microscopic origin of this effect.

\vskip 1.8cm

\section{Acknowledgments}
The single crystals were kindly supplied by N. Karl (Universit\"at
Stuttgart). The work in Italy was supported by the Ministero
Istruzione, Universit\`a e Ricerca (MIUR), through FIRB-RBNE01P4JF
and ~~PRIN2004033197\_002,~~ and in Stuttgart University by the
Deutsche Forschungsgemeinschaft (DFG). N.D. thanks the Alexander
von Humboldt foundation and the Scientific schools grant
NSH-5596.2006.2 for the support.

\end{document}